\documentclass[aps,prl,showpacs,twocolumn]{revtex4}
\usepackage{graphicx}

\begin{document}

\title{Theoretical evidence for a dense fluid precursor to crystallization}
\author{James F. Lutsko and Gr\'{e}goire Nicolis}
\affiliation{Center for Nonlinear Phenomena and Complex Systems, Universit\'{e} Libre de
Bruxelles, C.P. 231, Blvd. du Triomphe, 1050 Brussels, Belgium}
\date{\today}

\begin{abstract}
We present classical density functional theory calculations of the free
energy landscape for fluids below their triple point as a function of
density and crystallinity. We find that for both a model globular protein
and for a simple atomic fluid modeled with a Lennard-Jones interaction, it
is free-energetically easier to crystallize by passing through a metastable
dense fluid in accord with the Ostwald rule of stages but in contrast to 
the alternative of ordering and densifying at once as assumed in the
classical picture of crystallization. 
\end{abstract}

\maketitle

Crystallization is an intricate process of fundamental importance in many
areas of physics, chemistry and engineering. The classical picture of
crystallization from supersaturated solutions goes back to Gibbs and
consists of the spontaneous formation of crystalline clusters which then
either grow or shrink depending on the relative importance of the free
energy gain due to the lower bulk free energy of the crystal cluster and the
free energy penalty due to the surface tension between the two phases. In
this picture, the local density is the only order parameter: the crystalline
cluster is (in general) denser than the fluid. In recent years, this picture
has been called into question by simulation, theory and experiment for the
particular and important case of the crystallization of globular proteins.
ten Wolde and Frenkel (hereafter tWF) showed by means of simulation that the free energy
landscape of protein crystal clusters as a function of the number of atoms
in the cluster and the ''crystallinity '' favored paths leading from no
clusters to clusters with low order to ordered clusters over paths moving
from no clusters directly to ordered clusters\cite{tWF}. This picture was
confirmed by Talanquer and Oxtoby\cite{OxtobyProtein} and Shiryayev and
Gunton\cite{GuntonProtein} who showed using a parameterized van der
Waals-type model of globular proteins that surface wetting did indeed lower
the free energy of crystal clusters. More recently, the simple picture has
also been challenged by novel experimental investigations. Vekilov and
co-workers have shown that, prior to crystallization, protein solutions
harbor metastable droplets of dense fluid and they have suggested that these
droplets are necessary precursors of crystallization\cite%
{VekilovCGDReview2004,VekilovSpinodal2004,VekilovLusy2004,VekilovSpinodal2005}%
.\ The picture that emerges is one of the formation of metastable droplets
of dense fluid which then subsequently crystallize. In this paper, we show
by means of classical density functional theory calculations that there is
an \emph{intrinsic} free-energy advantage in first densifying into a
metastable dense-fluid state and then crystallizing rather than following
the classical path which goes directly from gas to crystal. Furthermore, our
calculations suggest that a similar advantage exists for fluids of small
molecules, modeled here via the Lennard-Jones (LJ) interaction, thus indicating
that this mechanism may underlie \emph{most }crystallization processes.

The starting point for our analysis is classical density functional theory
(DFT) which is based on a theorem, due to Mermin, that the Helmholtz free
energy of a classical system is a unique functional $F\left[ \rho \right] $
of the local density $\rho \left( \overrightarrow{r}\right) $. The local
density is a constant, $\rho \left( \overrightarrow{r}\right) =\overline{%
\rho }$ , for a bulk liquid while for a simple bulk solid it is a sum of
localized functions centered on the lattice sites,%
\begin{equation}
\rho \left( \overrightarrow{r}\right) =\sum_{i=0}f\left( \overrightarrow{r}%
-a_{0}\overrightarrow{R}_{i}\right) ,
\end{equation}%
for some function $f\left( \overrightarrow{r}\right) $where the vectors $%
\left\{ \overrightarrow{R}_{i}\right\} $ are the lattice vectors and the
lattice constant is $a_{0}$. Typically in a bulk solid, this is approximated
as a Gaussian, $f\left( \overrightarrow{r}\right) =\eta _{0}\left( \frac{%
\alpha }{\pi }\right) ^{3/2}\exp \left( -\alpha r^{2}\right) $, where $\eta
_{0}\in \left[ 0,1\right] $ is the fraction of lattice sites which are
occupied and the parameter $\alpha $ is related to the degree of
crystallinity. The average density for a lattice with $N_{0}$ lattice sites
per unit cell is $\overline{\rho }=\frac{1}{V}\int_{V}\rho \left( 
\overrightarrow{r}\right) d\overrightarrow{r}=\eta _{0}N_{0}a^{-3}$ where $V$
is the volume of the system. Given the Gaussian approximation the density
can also be written in terms of Fourier components as 
\begin{equation}
\rho \left( \overrightarrow{r}\right) =\overline{\rho }+\overline{\rho }%
\sum_{i=1}\exp \left( i\overrightarrow{K}_{i}\cdot \overrightarrow{r}%
/a_{0}\right) \exp \left( -K_{i}^{2}/4a_{0}^{2}\alpha \right) ,
\end{equation}%
where $\left\{ \overrightarrow{K}_{i}\right\} $ are the reciprocal lattice
vectors . This form shows clearly that as $\alpha $ goes to zero, the
density becomes uniform corresponding to a fluid whereas the real-space form
shows that as $\alpha $ goes to infinity, the density becomes infinitely
localized as a sum of Dirac delta functions. For this reason, it is natural
to take $\chi \equiv \exp \left( -K_{1}^{2}/4\alpha \right) $ to be an order
parameter corresponding measuring ''crystallinity '' since it becomes zero
for the liquid and one for the infinitely localized solid. The use of two
order parameters, average density $\overline{\rho }$ and crystallinity $\chi 
$, will allow us to explore different pathways from the gas/liquid to the
solid. Using two order parameters thus provides a richer space of possible
behaviors and intermediate states than does a single order parameter\cite%
{NicolisPhysica}.

In order to use DFT in practical calculations, it is of course necessary to
know the functional $F\left[ \rho \right] $. Good approximations exist for
this functional for the special case of hard-sphere interactions but the
extension of these to other potentials has proven difficult. For this
reason, liquid- and solid-state perturbation theory are often used as a
means of using the hard-sphere theory to approximate the functional for
other systems\cite{CurtinAschroftPert}. Indeed, simple manipulations yield
the exact relation%
\begin{widetext}
\begin{eqnarray}
\beta F\left[ \rho \right] &=&\beta F^{HS}\left[ \rho ;d\right] +\beta
\Delta F\left( \overline{\rho }\right) \\
&&-\int_{V}\int_{V}\left( \rho \left( \overrightarrow{r}_{1}\right) -%
\overline{\rho }\right) \left( \rho \left( \overrightarrow{r}_{2}\right) -%
\overline{\rho }\right) \int_{0}^{1}\left( 1-\lambda \right) \Delta
c_{2}\left( \overrightarrow{r}_{1},\overrightarrow{r}_{2};\left[ \rho
_{\lambda }\right] ,d\left[ \rho _{\lambda }\right] \right) d\lambda d%
\overrightarrow{r}_{2}d\overrightarrow{r}_{1}.  \nonumber
\end{eqnarray}%
\end{widetext}
where $\beta $ is the inverse temperature, $F^{HS}\left[ \rho ;d\right] $ is
the free energy functional for a hard-sphere system with hard-sphere
diameter $d\left[ \rho \right] $, $\Delta F\left( \overline{\rho }\right) $
is the difference in free energy of a liquid at density $\overline{\rho }$
and that of a hard-sphere liquid at the same density. In the integral, $%
\Delta c_{2}\left( \overrightarrow{r}_{1},\overrightarrow{r}_{2};\left[ \rho
_{\lambda }\right] ,d\left[ \rho _{\lambda }\right] \right) $ is the
difference in 2-body direct correlation functions (DCFs) for the interacting
system and the hard-sphere system for a density $\rho _{\lambda }\left( 
\overrightarrow{r}\right) =\overline{\rho }+\lambda \left( \rho \left( 
\overrightarrow{r}\right) -\overline{\rho }\right) $. The DCFs are not known
exactly and so it is necessary to introduce approximations to proceed.
Motivated by the fact that in applications of thermodynamic perturbation
theory to simple fluids, the correction to the hard-sphere free energy is
typically similar for FCC solids and liquids, we will make the simplest
approximation which is to assume that the contribution of the third term is
insignificant. This model has been shown to work well for the LJ potential\cite{Kyrlidis1993}
while using the more detailed model of Curtin and Ashcroft\cite%
{CurtinAschroftPert}, which is closely tied to the LJ potential,
we indeed find the third term to contribute little. More sophisticated
approximations will be discussed in a future publication. Here, our interest is not the further
development of DFT but in its application to the question of the kinetics of crystallization.

In the following, we use the first order WCA perturbation theory\cite%
{WCA1,WCA2,WCA3,HansenMcdonald} as modified by Ree et al\cite%
{Ree1,Ree2,ReeSol} to calculate the free energy of the liquid phase. This
theory is known to be very accurate for a wide class of potentials. The
liquid-phase hard-sphere diameter calculated from this theory is used for
both the liquid and the solid phases so that it is indeed solely a function
of the average density. For the hard-sphere free energy functional we use
the fundamental measure theory (FMT), specifically the ''White Bear ''
functional\cite{WhiteBear} which gives a good description of the hard-sphere
phase diagram, in particular reproducing the Carnahan-Starling equation of
state for the hard-sphere liquid. The use of the FMT free energy model is
critical:\ previous attempts to perform similar studies made use of
effective liquid approximations which do not work well when the occupancy is
treated as a free variable and which therefore had to be modified in an ad
hoc manner\cite{OLW,OxtobyBccFcc,OxtobyLJGrowth}. The FMT have built into
them the critical feature that they are sensitive to the local density and
correctly cause the free energy to diverge if the local occupancy grows
above one.

\begin{figure}[tbp]
\begin{center}
\resizebox{8.7cm}{!}{
{\includegraphics{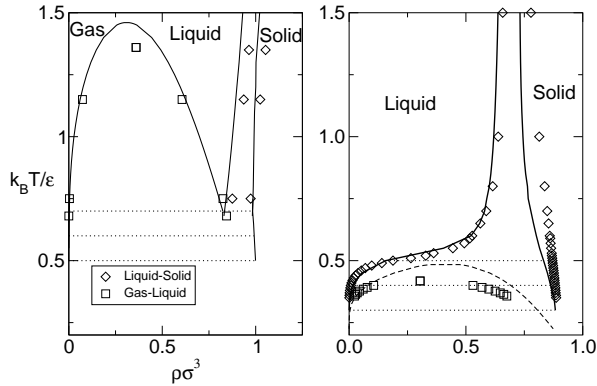}}}
\end{center}
\caption{The phase diagram for a LJ potential, left, and the ten
Wolde-Frenkel potential, right. The full and dashed lines are from the model and the points are from simulation, ref.\cite{VerletLJPhaseDiagram,HansenLJPhaseDiagram} and \cite{tWF} respectively. The dotted lines connect the coexistence points used in Figs. 2 and 3.}
\label{fig1}
\end{figure}

The calculations presented here were performed using the standard
LJ potential $v_{LJ}\left( r\right) =4\varepsilon \left( \left( 
\frac{\sigma }{r}\right) ^{12}-\left( \frac{\sigma }{r}\right) ^{6}\right) $%
, widely used as a model for the interactions of atomic fluids, and the
potential of tWF $v_{tWF}\left( r\right) $ which is
intended to model the interactions of globular proteins\cite{tWF}. The
latter consists of a hard-core of diameter $\sigma $ and a modified
LJ tail $v_{tWF}\left( r\right) =v_{LJ}(\lambda
^{1/6}(r^{2}-\sigma ^{2})^{1/2})$ for $r>\sigma $. where $\lambda $ controls
the range of the interaction; following ref. \cite{tWF}, we take $\lambda =50
$ . Figure \ref{fig1} shows the phase diagrams for both interaction models calculated
using our simplified DFT. (Note that the DFT also predicts a spinodal line but, for clarity, it has not been shown.) In both cases, the phase diagrams are reproduced
surprisingly well given the simple models used. The observed deviations from
the simulation data can be at least partly explained as arising from the use
of the liquid-state free energy difference for the solid which requires
knowledge of the liquid state at high densities for which even the input
hard-sphere equation of state is not reliable. Furthermore, the
determination of phase coexistence is a very sensitive test since it depends
on getting both the absolute magnitude and the slope of the free energies
correct. To put this in perspective, deviations are observed in Fig.\ref{fig1}
between the calculated and simulated gas-liquid coexistence curve for the
LJ system even though the liquid-state perturbation theory gives
free energies which differ from simulation by less than 1\% \cite{Ree2}. 

The difference between the phase diagrams of the two interaction models is
significant and generic. Whereas the LJ interaction gives rise to a typical
phase diagram with a critical point and, at lower temperatures, distinct gas
and liquid phases and a triple point, the tWF interaction model gives only a
single liquidus phase which is typical of short ranged interactions. Indeed,
as the parameter $\lambda $ in the tWF potential is varied from $\lambda =1$
to $\lambda =50$ , the phase diagram evolves continuously from one similar
to the LJ phase diagram to that shown here possessing a metastable
gas-liquid transition\cite{LutskoNicolis}. This model is motivated in part
by the fact that a dense metastable liquid phase is in fact experimentally
observed for some proteins. It is this metastable phase which tWF
showed to play a role in nucleation of the solid phase from the gas.

\begin{figure}[tbp]
\begin{center}
\resizebox{8.7cm}{!}{
{\includegraphics{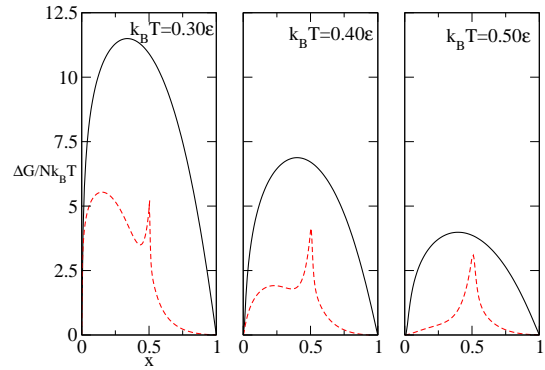}}}
\end{center}
\caption{The Gibbs free energy barriers, per atom, for the tWF potential. The solid curve is for the classical path and the broken curve results from densification followed by ordering.}
\label{fig2}
\end{figure}

\begin{figure}[tbp]
\begin{center}
\resizebox{8.7cm}{!}{
{\includegraphics{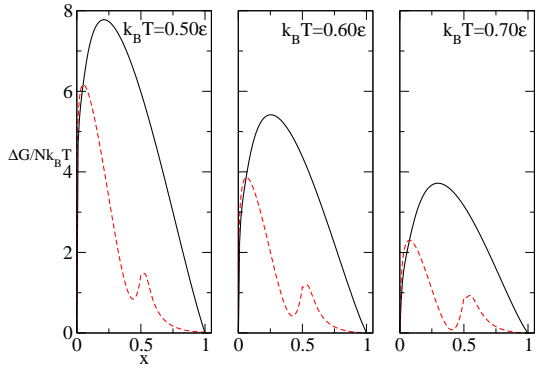}}}
\end{center}
\caption{Same as Fig.2 for the LJ potential.}
\label{fig3}
\end{figure}

Given a reasonable model for the DFT free energy functional, we now turn to
the question of the effect of different paths through density space from a
gas of density $\overline{\rho }_{gas}$ and crystallinity $\chi _{gas}=0$ to
a solid with density $\overline{\rho }_{solid}$ and crystallinity $\chi
_{solid}$. Here, we consider two candidate paths. The first corresponds to a
simultaneous densification and ordering of the gas into a solid and is
parameterized as 
\begin{eqnarray}
\overline{\rho }\left( x\right)  &=&\overline{\rho }_{gas}+x\left( \overline{%
\rho }_{solid}-\overline{\rho }_{gas}\right)  \\
\chi \left( x\right)  &=&x\chi _{solid}  \nonumber
\end{eqnarray}%
where $x\in \left[ 0,1\right] $ is an abstract reaction coordinate. This
might be thought of as the ''classical '' path. The second path we consider
is a two step process consisting of first a densification at zero
crystallinity followed by an ordering at fixed density 
\begin{eqnarray}
\overline{\rho }\left( x\right)  &=&\left[ \overline{\rho }_{gas}+2x\left( 
\overline{\rho }_{solid}-\overline{\rho }_{gas}\right) \right] \Theta \left( 
\frac{1}{2}-x\right)  \\
&&+\overline{\rho }_{solid}\Theta \left( x-\frac{1}{2}\right)   \nonumber \\
\chi \left( x\right)  &=&\Theta \left( x-\frac{1}{2}\right) \left(
2x-1\right) \chi _{solid}.  \nonumber
\end{eqnarray}%
Figure \ref{fig2} shows the free energy landscapes encountered using the tWF
potential along both paths for coexisting gas and solid densities at three
different temperatures. In all three cases, the classical path requires
overcoming a free energy barrier as expected. The behavior along the
non-classical path is more complex. At the highest temperature, which lies
about the critical point of the metastable gas-liquid transition, the
non-classical barrier is somewhat reduced but the effect is not significant.
For the intermediate temperature, which is somewhat below the critical
point, a second, metastable state appears and the single free energy barrier
splits into two lower barriers. At the lowest temperature, the barriers
encountered along the non-classical path are even lower so the advantage of
this path is even greater. This picture agrees well with that developed by
Vekilov and co-workers who have observed, by means of dynamic
light-scattering, the presence of short-lived dense liquid droplets in
protein solutions\cite{VekilovLusy2004}. 
The results for the LJ system, shown in Fig. \ref{fig3}, are unexpected in
that a similar phenomenon is observed although the details differ. Again, we
show three temperatures which are this time all below both the critical
point and the triple point. The highest temperature is only just below the
triple point and again, it is clear that the non-classical path is
energetically favored relative to the classical path and that this
correlates with the presence of a metastable dense-liquid state. Unlike the
previous case, the advantage of passing along the non-classical path remains
more or less constant as the temperature is decreased. Also different is the
fact that the barrier between the metastable state and the solid state is
much lower than that between the gas and the metastable liquid. This
suggests that the droplets in the metastable state will crystallize quickly
and will be correspondingly shorter-lived. To check this surprising result,
we repeated our calculations using the model described in ref.\cite%
{CurtinAschroftPert},\cite{OLW} which is tuned to the LJ
potential. While the barriers were somewhat smaller, the qualitative results
were the same. 

We have presented calculations of the free energy landscape as a function of
density and crystallinity based on a simple, robust free energy density
functional. Our calculations involve no input or parameterization except for
the interaction potential. In both cases studied, a model protein and a simple
liquid, our results provide direct support for the Ostwald rule of stages for nucleation\cite{OstwaldRef} since the
free energy barriers associated with the metastable intermediate states are lower
than those for a direct transition from gas to solid. The results for the model protein interaction
agree with the generally accepted picture that crystallization proceeds via
a two-step process of densification followed by crystallization\cite%
{VekilovCGDReview2004}, even when the temperature is slightly above the critical point
and no metastable intermediate phase exists. Interestingly, we find similar behavior for simple
fluids below the triple point suggesting that crystallization involving passage through a metastable disordered state may be a generic phenomenon. However, in that case, the metastable state
is expected to be shorter-lived compared to the nucleation time thus making
its experimental detection more challenging. The only evidence we are aware
of for the two-step nucleation mechanism for non-protein fluids comes from
nonphotochemical laser-induced nucleation of small organic molecules\cite%
{GaretzLaser1,GaretzPrecursorPolarization,OxtobyPrecursor} and recent
molecular dynamics simulations of the crystallization of AgBr from solution%
\cite{MDNucleation}. The short lifetime of the metastable phase predicted
here would explain why it has not so far been observed experimentally in
simple fluids.

For the protein model, our results indicate similar barriers for the
gas-liquid and liquid-solid transitions and it should be noted that the only
experimental results indicate that the latter should be much higher than the
former\cite{VekilovSpinodal2005}. In part, this is because we have presented
results for the free energy landscape for transitions near the coexistence
lines. For denser gases, which are supersaturated with respect to
crystallization, the free energy of the gas phase moves up so the first
barrier is smaller. However, it is important to notice that the crossing of
the free energy barriers is a fundamentally non-equilibrium process so that
kinematics also plays an important role in determining the overall
nucleation rates\cite{VekilovCGDReview2004},\cite{NicolisPhysica}. Nevertheless, if crystallization kinetics occurs reasonably close to equilibrium the free energy functional will play a central role since the rates of change of the order parameters will be given by the product of its gradient and of a matrix of phenomonological parameters.

A point which could cause concern is that the paths through parameter space might  cross the spinodal and so  
pass through the two-phase region where it could be thought that the use of DFT is problematic. In fact, aside from the minima, \emph{all} points on the curves shown in Fig. 2 are thermodynamically unstable. However, the fundamental 
idea underlying DFT is that any density profile can be stabilized by means of an external field and that the free energies
calculated are the intrinsic contribution to the free energy when such a stabilizing field is present (see, e.g. ref. \cite%
{HansenMcdonald}). Only the intrinsic contribution is used here to estimate the barriers as the real system must pass through these states without the presence of such a stabilizing field.

One question not answered yet is whether the particular pathways discussed
here are the optimal, i.e. minimum energy, pathways. Simple contour plots of
the free energy show that the non-classical paths used here are indeed very
close to the optimal paths as will be discussed at length in a future
publication. Another question is the role of surface tension which should in
general increase the free energy barriers, as well as the free energies of
clusters in the metastable and solid states. Since the penalty due to
surface tension is expected to increase as the system passes from the gas to
the metastable state to the solid state, we expect that the barriers and the
free energy minima will be shifted accordingly but it seems unlikely that
the overall picture would change since this would require that the addition
of surface tension affect the classical path less than the non-classical
path. A definitive answer to this question will require calculations of free
energies for inhomogeneous states which we are currently pursuing.

\begin{acknowledgements}
It is our pleasure to thank Pieter ten Wolde and
Daan Frenkel for making their simulation results available to us.
We benefited greatly from discussions with Peter Vekilov and Bruce Garetz.
This work was supportd in part by the European Space Agency
under contract number C90105.
\end{acknowledgements}

\bibliographystyle{prsty}
\bibliography{physics}

\end{document}